\documentclass[prl,twocolumn,showpacs,preprintnumbers,amsmath,amssymb]{revtex4}
\usepackage{graphicx}% Include figure files
\usepackage{dcolumn}% Align table columns on decimal point
\usepackage{bm}% bold math

\begin{document}

\preprint{PRL Preprint}

%\title{Pressure, surface tension and capillarity of light condensates}
               
\title{Dripping, pressure and surface tension of self-trapped laser beams}
\author{David Novoa, Humberto Michinel and Daniele Tommasini}
\affiliation{Departamento de F\'{\i}sica Aplicada, Facultade de Ciencias de Ourense,\\
             Universidade de Vigo, As Lagoas s/n, Ourense, ES-32004 Spain.}

\begin{abstract}
%----------------------------   ABSTRACT  ------------------------------------
We show that a laser beam which propagates through an optical medium with Kerr (focusing)
and higher order (defocusing) nonlinearities displays pressure and surface-tension properties 
yielding capillarity and dripping effects totally analogous to usual liquid droplets.
The system is reinterpreted in terms of a thermodynamic grand potential, allowing for the 
computation of the pressure and surface tension beyond the usual hydrodynamical 
approach based on Madelung transformation and the analogy with the Euler equation.
We then show both analytically and numerically that the stationary soliton states of such a 
light system satisfy the Young-Laplace equation, and that the dynamical evolution through a 
capillary is described by the same law that governs the growth of droplets in an ordinary 
liquid system.
\end{abstract}

\pacs{03.75.Lm, 42.65.Jx, 42.65.Tg}

\maketitle

%---------------- INTRODUCTION --------------------

{\em Introduction.-}
Since the pioneering paper of Piekara in the 70's\cite{piekara74}, many works have highlighted 
the interesting properties of laser beams whose propagation is described by the so-called 
{\em cubic-quintic} (CQ) nonlinear Schr\"odinger equation (NLSE) with competing nonlinearities.
Cavitation, superfluidity and coalescence have been investigated\cite{josserand1,josserand2} in the context 
of liquid He, where the model is a simple approach which does not take into account nonlocal
interactions. Stable optical vortex solitons and the existence of top-flat states have also been
reported in optical materials with CQ optical susceptibility\cite{quiroga97}. Recent experiments 
about filamentation of high-power laser pulses have shown that the CQ regime could be achievable 
in $CS_2$ \cite{centurion05} as well in some chalcogenide glasses\cite{smektala00}. Recently, it 
has been also suggested that atomic coherence may be used to induce a giant CQ-like refractive
index in Rb gas \cite{michinel06}.

On the other hand, this model has been shown to display surface properties in numerical simulations 
of soliton collisions \cite{edmundson95}, that have been considered as a trace of a liquid state 
of light\cite{michinel02,michinel06}. Here, we will provide the first analytical, quantitative
demonstration of the liquid behavior of the system in ($2+1$) dimensions, 
both in its stationary soliton solutions and in 
the dynamical evolution when a light bump is forced to pass through a wave-guide simulating 
a capillary. In particular, we will provide the first consistent computation of the 
pressure and surface tension of the light bubbles in ($2+1$) dimensions,
and show both analytically and numerically that they satisfy the Young-Laplace (Y-L) 
equation that gives the equilibrium of usual droplets. Subsequently, we will demonstrate that the 
system dripping properties are governed by the same generalized Y-L that applies to an ordinary 
liquid. These results show the deep connection between the nonlinear dynamics of laser 
beams and coherent liquids at zero temperature \cite{Chiao}.

%-----------THERMODYNAMIC MODEL--------------

{\em Thermodynamic model.-} We will consider the paraxial propagation through an ideal CQ medium
of a linearly polarized laser beam, being its complex amplitude distribution $\Psi$ described by a 
nonlinear Schr\"odinger equation of the adimensional form:
\begin{equation}
\label{NLSE}
i\frac{\partial \Psi }{\partial z}+\frac{1}{2}\nabla _{\perp }^{2}\Psi
+\gamma|\Psi |^{2}\Psi-\delta|\Psi |^{4}\Psi=0,
\end{equation}
where $z$ is the propagation distance multiplied by $2\pi/\lambda$, being $\lambda$ 
the wavelength of the continuous light beam; $\nabla_{\perp}^{2}$ is the transverse Laplace 
operator in terms of $x,y$, the spatial variables multiplied by $2\pi\sqrt{2n_{0}}/\lambda$,
being $n_{0}$ the linear refractive index of the medium; $\gamma$ and $\delta$ are proportional
to the (opposite) $\chi^{(3)}$ and $\chi^{(5)}$ optical susceptibilities, respectively.

It is well-known that stationary version of Eq.\eqref{NLSE} admits localized soliton-like 
solutions \cite{piekara74} of the form $\Psi_A(x,y,z)=A(x,y)e^{-i\mu z}$, being $\mu$ 
the propagation constant. In particular, it has been shown numerically that high power solitons 
feature top-flat profiles\cite{quiroga97}. These modes can only be calculated numerically 
and coexist with plane waves solutions of constant amplitude $\Psi_A(x,y,z)=A e^{-i\mu z}$, 
which lead by substitution in Eq.\eqref{NLSE} to $\mu=\delta\vert A\vert^4-\gamma\vert A\vert^2$.
The existence domain for solitons is $\mu\in(\mu_{\infty},0)$, where $\mu_{\infty}=-\frac{3\gamma^2}{16\delta}$.

As discussed in \cite{physicaD}, the stationary solutions of Eq. \eqref{NLSE} can be derived 
from a variational principle $\frac{\delta\Omega}{\delta\psi^*}=0$ from the Landau's 
grand potential $\Omega=H-\mu N$, where $H$ is the Hamiltonian, $N=\int|\Psi|^2d{\cal S}$ 
the particle number (in our system the photon flux through the transverse section ${\cal S}$) 
and $\mu$ the chemical potential which is thus identified in our case as the propagation constant 
we have defined above. The resulting expression is:

\begin{eqnarray}
\label{Omega}
\Omega&=&\int \left[\frac{1}{2}\vert\nabla _{\perp }\Psi\vert^{2}
-\frac{\gamma}{2}|\Psi |^{4}+\frac{\delta}{3} |\Psi |^{6} -\mu |\Psi |^{2}\right]d{\cal S},
\end{eqnarray}

In three dimensional systems, the partial derivative $\left(\partial\Omega/\partial { V}\right)_{\mu,T}$
of $\Omega$ with respect to the volume $V$ at constant chemical potential $\mu$ and temperature $T$
would give minus the pressure. Assuming a completely coherent two-dimensional model with $T=0K$, it 
is then natural to use the derivative of $\Omega$ with respect to the area $\cal S$, which yields:
\begin{equation}
p=-\left(\partial\Omega/\partial {\cal S}\right)_{\mu}=-\frac{1}{2}\vert\nabla _{\perp }\Psi\vert^{2}
+\frac{\gamma}{2}|\Psi |^{4}-\frac{\delta}{3} |\Psi |^{6} +\mu |\Psi |^{2}.
\label{pressure_general}
\end{equation}

%********************** fig 1  ******************************
\begin{figure}[htbp]
{\centering \resizebox*{1\columnwidth}{!}{\includegraphics{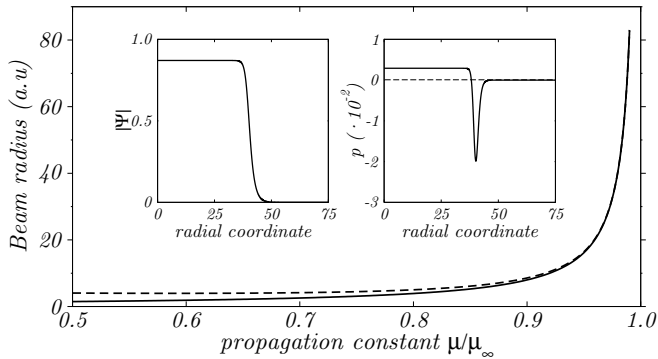}} \par}
\caption{[Color online]
Plot of radius vs. propagation constant for different localized stationary solutions of Eq.\eqref{NLSE}.
Solid (dashed) line corresponds to numerical (analytical) calculations. The left (right) inner picture, 
shows the field modulus profile (pressure distribution) of a stationary beam with $\mu/\mu_{\infty}=0.98$.
Notice that close to the origin, the homogeneous pressure is positive whereas in the inhomogeneous region
becomes negative.}
\label{fig1}
\end{figure}
%***************************************************************************

In Fig.\ref{fig1} we plot an example of a "flat-top" stationary state, corresponding to $\mu/\mu_{\infty}=0.98$, 
together with its pressure distribution. For all this type of solutions, in the wide flat region
the value of the pressure is positive and equal to the central value $p(0)\equiv p_c$.  Close to the origin 
it is straightforwardly obtained for the beam amplitude:
\begin{equation}
|A(0,0)|^{2}=\frac{\gamma }{2\delta }+\frac{\sqrt{\gamma ^2+4 \delta  \mu }}{2 \delta }
\label{A_mu}
\end{equation}
which yields for the pressure at the center of the beam ($p_c$) the following analytical expression:
\begin{equation}
p_c=\frac{(\gamma + \sqrt{\gamma^2   + 4 \delta\mu} )^2  
(-\gamma  + 2 \sqrt{\gamma^2  + 4 \delta \mu })}{24 \delta^2 }.
\label{pressure_pc}
\end{equation}
Notice that $p_c$ vanishes at the limits of the existence domain: $\mu=0$ (trivial zero-amplitude solution)
and $\mu=\mu_{\infty}$ (infinite "flat-top" with $|A_{\infty}|^2=0.75\gamma/\delta$). This result
can be considered as a trace of the existence of two possible vacuum states in the system\cite{physicaD}.

%---------- Limitations of the hydrodynamical analogy------------------
{\em Limitations of the hydrodynamical analogy.-} Our present purpose is to use the previous formalism 
in order to study the physical properties of the system. First, let us try to apply the commonly used 
hydrodynamical analogy, based on introducing the so-called Madelung Transformation\cite{Madelung}
$\psi=\rho^{1/2} e^{i\phi}$, where $\rho$ is a positive definite real function and $\phi$ is a real phase, 
eventually depending on the variables $x$, $y$ and $z$. After substituting in Eq.\eqref{NLSE}, and 
separating both the real and the imaginary part, we get two equations. The first one is a continuity equation, 
that is used to establish an analogy with hydrodynamics, and to argue that $\nabla \phi$ can be identified 
with a current, i.e. a velocity field ${\bf v}$. The second equation is

\begin{equation}
\frac{\partial{\bf v}}{\partial z}+\nabla_{\perp } \left(\frac{{\bf v}^2}{2}
+\frac{1}{2\rho^{1/2}}\nabla_{\perp }^2\rho^{1/2}+\gamma\rho-\delta\rho^{2}\right)=0.
\label{NLSE_euler}
\end{equation}

Here, the usual approach\cite{josserand1} is to assume 
that the term $\frac{1}{2\rho^{1/2}}\nabla_{\perp }^2\rho^{1/2}$
can be neglected. In this case, Eq. \eqref{NLSE_euler} is identical to the known Euler equation of hydrodynamics
provided that $\nabla_{\perp } (\gamma\rho-\delta\rho^{2})=-\frac{\nabla_{\perp } p}{\rho}$.
This formula was used e.g. in Ref. \cite{josserand1} in order to derive an expression 
for the pressure of the homogenous phase, which coincides
with our result of Eq. \eqref{pressure_pc} for the top flat region of the solitons.
One could hope that this analogy could be generalized also to the non-homogenous zones, such as 
that corresponding to values of the radial coordinate $r$ around the radius $R$ of the beam.
However, we will see that this is not the case. For a (non-rotational) top flat soliton the phase term
is simply equal to $\phi=-\mu z$, therefore ${\bf v}=\nabla_{\perp } \phi=0$. Note that $\mu$ is the 
propagation constant which is just a constant number for the given soliton. Therefore, by substituting in 
Eq. (\ref{NLSE_euler}), we get
\begin{equation}
\nabla_{\perp} \left(\frac{1}{2\rho^{1/2}}\nabla_{\perp}^2\rho^{1/2}+\gamma\rho-\delta\rho^{2}\right)=0.
\label{noeuler_soliton}
\end{equation}

In other words, in this case the term that is usually neglected
is {\it exactly} equal and opposite to the part that is used to compute the pressure.
Therefore, neglecting such a term would correspond to a 100\% error. In fact, for any soliton 
solution, having ${\bf v}=0$, Euler equation would unavoidably imply a constant pressure, which 
cannot be the case in any region where the spatial variation of the density is important. Nevertheless, 
in spite of this failure of the ordinary hydrodynamical approach, we will see that the 'pressure' 
distribution in the non-homogeneous region can still be given a deep physical interpretation.

%--------- Equilibrium and surface tension.------------------

{\em Equilibrium and surface tension.-}
Fig. 1 suggests that the $\Omega$ potential, as given by the spatial integral
of minus the pressure, can be expressed as the sum of two contributions:
one from the top-flat region, which is $-\pi R^2 p_c$ with a good accuracy; and
another from the region near the border, where the field is spatially-dependent,
which is $-2\pi\int_R^\infty{r p(r)dr}$. Thus, we get the following analytical 
approximation for the $\Omega$ potential of the "flat-top" solutions:
\begin{equation}
\Omega\simeq
-\pi R^2 p_c+2\pi \sigma R,
\label{Omega_app}
\end{equation}
where we have defined a parameter $\sigma\equiv -\frac{1}{R} \int_R^\infty{r p(r)dr}$.
We will now argue on how $\sigma$ can be identified with the surface tension of the 
light beam. In first place, we have computed numerically the parameter $\sigma$
for the different high-power top-flat solitons, and we have found that it converges
quickly to a fixed value $\sigma\simeq 0.057$ as soon as $\vert\mu\vert$ approaches 
the limiting value $\mu_{\infty}$. In such a limit, the central pressure goes to zero 
and the only important contribution to the integral defining $\sigma$ comes
from the gradient term of Eq. \eqref{pressure_general}.
Thus $\sigma\simeq -\frac{1}{2 R} \int_R^\infty{r\vert\nabla _{\perp }\Psi\vert^{2}dr}$.

We can then get an analytical approximation for $\sigma$ by noting that for large $R$ and $r$, 
$\nabla _{\perp }\Psi\simeq d\Psi/dr$, and the multiplicating $r$ in the integrand can be approximated by $R$ in the 
comparatively thin 'surface' region. Taking into account that for large $r$ 
Eq. \eqref{NLSE} yields $A'(r)^{2}\simeq-\gamma A(r)^{4}/2+\delta A(r)^{6}/3+\mu A(r)^{2}$,
and by substituting the limiting value $\mu\to\mu_{\infty}=-\frac{3\gamma^{2}}{16\delta}$,
we get after some algebra the expression:
\begin{equation}
\sigma\simeq
\frac{9\gamma^{2}}{64\sqrt{6}\delta^{3/2}}.
\label{sigma_limit}
\end{equation}
For instance, the choice $\gamma=\delta=1$ gives $\sigma=0.057$,
in complete agreement with our numerical value. By differentiating Eq. \eqref{Omega_app} 
and taking into account that $\sigma$ is constant, we get:
\begin{equation}
\label{d_Omega_app}
\frac{d\Omega}{d R}\simeq
-\pi\left[\frac{d (R^2 p_c)}{dR}-2\sigma\right].
\end{equation}
Our numerical calculations show that the growth in $R$ holds the thermodynamical equilibrium, 
(i.e.: $\frac{d\Omega}{d R}\simeq0$), within a relative error which turns out to be as small as
$\frac{d\Omega}{d R}<10^{-4}\frac{\Omega}{R}$ for all the range of "flat-top" eigenmodes considered.
Therefore, we can set $\frac{d\Omega}{d R}=0$ in Eq. \eqref{d_Omega_app}, and we get 
$\frac{d (R^2 p_c)}{dR}=2\sigma$, or
\begin{equation}
\label{pressuretoradius}
p_c=2\frac{\sigma}{R},
\end{equation}
which is the celebrated \emph{Young-Laplace} (Y-L) equation\cite{Landau} for spherical liquid 
droplets being $\sigma$ the surface tension of the system. Therefore, despite the failure of
the usual hydrodynamic approach, we have demonstrated that the pressure distribution in the inhomogeneous 
region has a deep physical interpretation. In fact, the integral of the pressure in the non-homogeneous 
surface region gives the surface tension ($\sigma$), i.e. the inward force that compensates the outward 
positive inner force described by the pressure $p_c$, in order to keep the droplet stationary.

%********************** fig 2  ******************************
\begin{figure}[htbp]
{\centering \resizebox*{1\columnwidth}{!}{\includegraphics{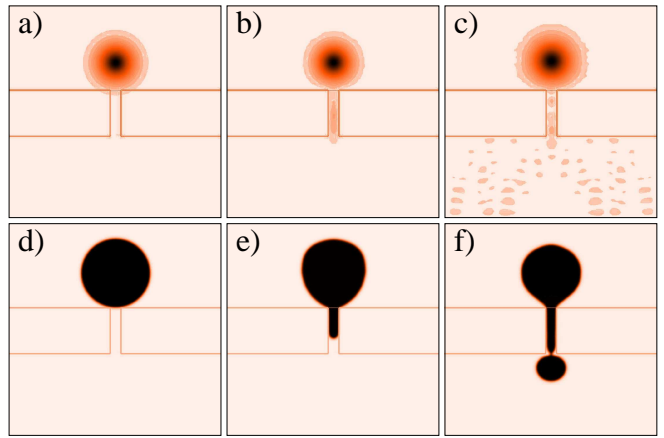}} \par}
\caption{[Color online]
Numerical simulation of the dripping behavior of light beams propagating through a channel waveguide 
highlighted with solid lines. As a comparison, the upper (lower) row corresponds to snapshots showing 
the evolution of the modulus of the quasi-gaussian ("flat-top") solutions. The size of the pictures in
x and y is the interval $ [-250,250]$. The snapshots a)-d), b)-e), f) and c) correspond to propagation
distances $z=0,11000,27000,45000$, respectively.}
\label{fig2}
\end{figure}
%***************************************************************************

Instead of directly comparing Eq.\eqref{pressuretoradius} with the numerical simulation, we will equivalently test
the inverted equation $R=2\sigma/p_c$, where $p_c$ can be expressed analytically in terms either of the central
amplitude $A$ or of the propagation constant $\mu$, by using Eq. \eqref{pressure_pc} and
Eq.\eqref{sigma_limit}. This result provides the first analytical expression
for the radius $R$ of the bidimensional 
top-flat solitons as a function e.g. of $\mu$, and is compared with the  
numerical solutions of the stationary version of Eq.\eqref{NLSE} in Fig.\ref{fig1}. As it can be 
appreciated in the figure, the agreement between the analytical formula and the numerical computation 
is remarkable, and becomes complete when $|\mu|$ approaches $|\mu_{\infty}|$. This result confirms our 
theoretical framework and provides the first formal demonstration, by validation of the Y-L equation,
of the liquid properties of the "flat-top" solutions in the ($2+1$) dimensional CQ model. 
Moreover, from Eq.\eqref{pressuretoradius} 
it can be inferred that, as $R\rightarrow\infty$, the value of $p_{c}$ vanishes, indicating that surface tension
effects are not needed to balance the inner pressure, as it is the case of standard liquids described 
by the Y-L equation.

%-------  Dripping of light droplets.------------------

\begin{figure}[htbp]
{\centering \resizebox*{1\columnwidth}{!}{\includegraphics{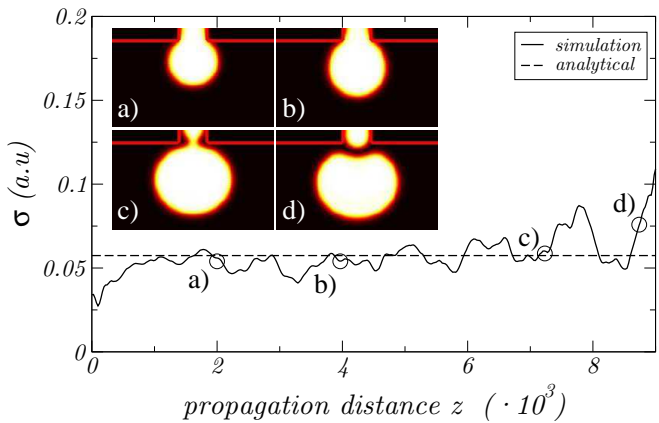}} \par}
\caption{[Color online]
Numerically calculated values of the surface tension along the droplet-formation process.
Solid (dashed) line corresponds to numerical (analytical) calculations of $\sigma$.
Insets: snapshots showing different stages of evolution of the light droplet for different values of $z$.
Circles in the figure refer to upper insets. The spatial scales spanned are $x\in[-50,50],y\in[-20,-80]$.}
\label{fig3}
\end{figure}

{\em Dripping of light droplets.-} In classical fluid mechanics, the presence of surface tension effects 
can be appreciated in the dynamical phenomenon of capillarity \cite{Landau,surften_review}.
Here, in order to study droplets formation and dripping in our system, we have introduced in 
our mathematical model an external ``channel-type'' linear optical waveguide $V(x,y)$, superposed to the cubic-quintic
nonlinearity. This waveguiding structure consists of three regions with indices $n_{1}=0.002$ $n_{2}=0.0028$ and
$n_{3}=0.001$, which correspond to the top region($n_{1}$), central channel and bottom zones($n_{2}$) and rectangular 
regions flanking the central channel($n_{3}$), respectively, as it can be seen in the pictures of Fig.\ref{fig2}.
In our simulations, we compare the evolution of two initial eigenstate beams with propagation constants  
$\mu/\mu_{\infty}=0.01$ (quasi-gaussian beam of Fig.\ref{fig2}a) and $\mu/\mu_{\infty}=0.98$ ("flat-top" beam 
of Fig.\ref{fig2}d), both located within the waveguide top region. Depending on the channel size, and keeping 
$n_3<n_2,n_{1}$, above a given value of $\Delta n= n_{2}-n_{1}$ in both cases a significant amount of beam power 
starts to flow from the initial eigenstate through the channel, as shown in (Figs.\ref{fig2}b and \ref{fig2}e).
It is noteworthy that the light stream inside the channel does not suffer any unstabilization and remains 
\emph{connected} to the initial source of light, i.e., the guide prevents the appearance of modulational instability.  
At the output of the channel, it can be seen in Fig.\ref{fig2}c that the low-power distribution spreads like a 
(coherent) gas in free expansion. However, the light flowing from the "flat-top" beam yields to the formation 
of a droplet, as it can be appreciated in Fig.\ref{fig2}f and in more detail in the insets of Fig.\ref{fig3}.

As it can be seen in the insets of Fig.\ref{fig3}, just before the release of the droplet the falling column 
of liquid light becomes narrower, just as in the well-known case of usual liquid streams falling from a 
tap\cite{dripping2}. On the other hand, all the steps in Fig.\ref{fig2} are qualitatively very similar to 
those obtained both in real experiments and in simulations with liquids \cite{surften_review}-\cite{dripping1}.

%********************** fig 3  ******************************
%***************************************************************************

Finally, we have also performed a {\it quantitative} test by comparing the numerical simulation with the generalized
Y-L equation that describes the growth of an elliptic bubble of an ordinary liquid system,
$p_c=\sigma(\frac{1}{R_I}+\frac{1}{R_{II}})$, where $R_I$ and $R_{II}$ are the principal radii
\cite{Landau}.
In Fig.\ref{fig3}, we have plotted the quantity $\sigma=p_c/(\frac{1}{R_I}+\frac{1}{R_{II}})$
for the ``dripping'' simulation of Fig.\ref{fig2} at several propagation distances, assuming $z=0$ where the light 
droplet first appears.
We see that the numerical value of $\sigma$ oscillates around the same value that we have calculated analytically in 
Eq.\eqref{sigma_limit} for the stationary solutions. This demonstrates that the droplets are formed
close to stationary equilibrium, and they grow according to the generalized 
Y-L equation as in the case of an ordinary liquid.

{\em Conclusions.-}.
We have provided the first consistent computation of the pressure and surface 
tension of the soliton solutions appearing in the propagation of 
self-trapped laser beams described by the ($2+1$) dimensional CQ-NLSE, and we have shown 
both analytically and numerically that they satisfy the 
Young-Laplace equation that governs the equilibrium of usual liquid droplets. 
Subsequently, we have also demonstrated that the system dripping properties are governed by 
the same generalized Y-L that applies to an ordinary liquid. These results reveal the
deep connection between the physics of self-trapped laser beams and quantum liquids at zero temperature,
opening the door for the quest of these new states of matter in the frame of current nonlinear optics
experiments.

{\em Acknowledgements.-} We thank A. Ferrando for useful
discussions. This work was supported by MEC, Spain (projects FIS2006-04190
and FIS2007-62560) and Xunta de Galicia (project PGIDIT04TIC383001PR
and D.N. grant from Conseller\'{i}a de Innovaci\'on e Industria-Xunta de Galicia).

%----------------------------------------------------------------------------

\end{document}